\documentclass[superscriptaddress,aps,twocolumn,showpacs,nofootinbib]{revtex4}

\usepackage{graphicx}
\usepackage{amsmath,amssymb}

\begin{document}

\title{Natural inflation in 5D warped backgrounds}

\author{R. Gonz\'{a}lez Felipe}
\email{gonzalez@cftp.ist.utl.pt}
\affiliation{%
Instituto Superior de Engenharia de Lisboa\\ Rua Conselheiro Em\'idio Navarro, 1959-007 Lisboa,
Portugal}
\affiliation{%
Centro de F\'{\i}sica Te\'orica de
Part\'{\i}culas, Instituto Superior T\'{e}cnico\\ Avenida Rovisco
Pais, 1049-001 Lisboa, Portugal}

\author{N.~M.~C. Santos}
\email{ncsantos@cftp.ist.utl.pt}
\affiliation{%
Centro de F\'{\i}sica Te\'orica de
Part\'{\i}culas, Instituto Superior T\'{e}cnico\\ Avenida Rovisco
Pais, 1049-001 Lisboa, Portugal}

\begin{abstract}

In light of the five-year data from the Wilkinson Microwave Anisotropy Probe (WMAP), we discuss models of inflation based on the pseudo Nambu-Goldstone potential predicted in five-dimensional gauge theories for different backgrounds: flat Minkowski, anti-de Sitter, and dilatonic spacetime. In this framework, the inflaton potential is naturally flat due to shift symmetries and the mass scales associated with it are related to 5D geometrical quantities.

\end{abstract}


\date{\today}
\maketitle

\section{Introduction}

Inflation~\cite{Guth:1980zm,Linde:1981mu,Albrecht:1982wi,Linde:1983gd} is at present the favorite paradigm for explaining both the cosmic microwave background radiation (CMBR) temperature anisotropies and the initial conditions for structure formation~\cite{Mukhanov:1981xt,Guth:1982ec,Hawking:1982cz,Linde:1982uu,Starobinsky:1982ee,Bardeen:1983qw,Lyth:1984gv}. In the simplest inflationary models, the energy density of the Universe is dominated by a single scalar field $\phi$, the inflaton, that slowly rolls down its self-interaction potential. The results of the Wilkinson Microwave Anisotropy Probe (WMAP)~\cite{Bennett:2003bz,Spergel:2003cb,Hinshaw:2006ia,Page:2006hz,Spergel:2006hy,Hinshaw:2008kr,Dunkley:2008ie,Komatsu:2008hk} support the standard inflationary predictions, i.e., the Universe is consistent with being flat and the primordial fluctuations are adiabatic, nearly scale-invariant and well described by random Gaussian fields~\cite{Komatsu:2003fd}.

To satisfy the observational constraints, the potential for the inflaton field must be sufficiently flat. This leads to the so-called "fine-tuning" problem in inflation, since couplings must be adjusted to keep the inflaton weakly self-coupled and the radiative corrections under control. Despite the existence of several cosmologically viable inflaton potentials, the construction of a ``naturally" flat potential is a difficult task from the particle physics viewpoint. Among the models of inflation which do not suffer from the above fine-tuning problem is the so-called natural inflation, based on nonlinearly realized symmetries. Particularly simple are those realizations which involve a pseudo Nambu-Goldstone boson (pNGB)~\cite{Freese:1990rb} or extra components of gauge fields propagating in extra dimensions~\cite{Arkani-Hamed:2003wu,Kaplan:2003aj,Arkani-Hamed:2003mz}.

In the simplest variant of natural inflation~\cite{Freese:1990rb}, the role of the inflaton is played by a pNGB, and the flatness of the potential is protected by a shift symmetry under $\phi \rightarrow \phi + {\rm constant}$, which remains after the global symmetry is spontaneously broken. An explicit breaking of the shift symmetry leads typically (in four spacetime dimensions) to a potential of the form
\begin{align}\label{pot}
V(\phi)= \Delta^4\, [1 \pm \cos(\phi/f)]\,,
\end{align}
where $\phi$ is the canonically normalized field, $f$ is the spontaneous breaking scale and $\Delta$ is the scale at which the soft explicit breaking takes place. For large values of $f$, the potential can be flat. In this framework, the slow-roll requirements set the bound $f \gtrsim M_P \equiv (8\pi G)^{-1/2} \simeq 2.4 \times 10^{18}$~GeV ($M_P$ is the 4D reduced Planck mass), while $\Delta\sim 10^{15-16}$~GeV produces the required amplitude of scalar perturbation measurements. Although from a field theoretical description it seems difficult to justify such Planckian values of the scale $f$, they can be generated from effective field theories. For instance, in the 5D version of natural inflation considered in Ref.~\cite{Arkani-Hamed:2003wu}, the inflaton is the extra component of a gauge field, which propagates in the bulk, and the flatness of its potential, coming only from nonlocal effects, is not spoiled by higher-dimensional operators or quantum gravity effects. In this model, and in further works~\cite{Kaplan:2003aj, Arkani-Hamed:2003mz, Fairbairn:2003yx, Feng:2003mk, Hofmann:2003ag, PaccettiCorreia:2005zm}, the bulk is taken to be flat.

In an extra-dimensional framework, the standard KK decomposition approach~\cite{Hosotani:1983xw} has been extensively applied to Higgs physics. This includes models of gauge-Higgs unification in flat space~\cite{Hatanaka:1998yp,Antoniadis:2001cv,Scrucca:2003ra,Panico:2005dh} and warped space~\cite{Agashe:2005dk,Falkowski:2006vi,Contino:2006qr}. More recently, an alternative approach based on holography has also been successfully applied to theories on the interval~\cite{Luty:2003vm, Contino:2003ve, Barbieri:2003pr,Agashe:2004rs, Panico:2007qd}. One of the difficulties encountered in deriving the 4D effective scalar potential from a higher-dimensional theory is the computation of form factors that typically describe interactions of the low-energy effective theory. In Ref.~\cite{Falkowski:2006vi}, a systematic method to compute the 4D pNGB effective potential and the corresponding form factors for an arbitrary 5D warped background has been presented, and the results applied to the phenomenology of the electroweak symmetry breaking and Higgs physics.  In particular, it is argued that the 4D effective pNGB potential can be well approximated by~\cite{Falkowski:2006vi}
\begin{align}
V(\phi) = \sum_r\frac{N_r}{(4\pi)^2}\!\! \int_0^\infty\!\! dp \,p^3
\ln\!\left[1+\frac{g_r^2 f^2}{M^2_{c}}\frac{\sin^2(\lambda_r\,
\phi/f)}{\sinh^2(p/M_{c})}\right],\label{eq:pGpot}
\end{align}
where $f$ is the scale of spontaneous global symmetry breaking, $g_r$ is an effective low-energy coupling, $N_r=+3\, (-4)$ for gauge (fermion) particles, and $\lambda_r$ is a discrete number that depends on the gauge group structure. The sum is performed over all (gauge and fermionic) zero modes. The Kaluza-Klein (KK) scale $M_{c}$ is roughly equal to the mass of the lightest KK mode and corresponds to the compositeness scale in 4D.

In this paper we investigate models of inflation driven by the pNGB potential of Eq.~(\ref{eq:pGpot}), as predicted in 5D gauge theories. As we shall see, this potential is nearly flat (cosine-type) for $g_r f \leq M_c\,$, and therefore, it is a suitable candidate for natural inflation. In particular, we shall consider natural inflation in different backgrounds (flat Minkowski, anti-de Sitter, or dilatonic spacetime). In this framework, it is possible to relate the inflationary mass scales to five-dimensional geometrical quantities. Furthermore, using the inflationary constraints coming from the recent WMAP five-year data (WMAP5), we can put constraints on the relevant mass scales. Gonz\'{a}lez

\section{Natural inflaton potential}

We consider a 5D gauge theory with a gauge group $\mathcal{G}$ and the fifth dimension compactified in a finite interval, $y \in [0,L]$, such that the UV and IR branes are located at $y=0$ and $y=L$, respectively. The four spacetime dimensions are assumed to be flat, while the gravitational background can be warped, with the line element given by:
\begin{align}
ds^2 = a^2(y)\, \eta_{\mu \nu}\, dx^\mu dx^\nu-dy^2~,
\end{align}
where $a(y)$ is the warp factor, $a(0)=1$ and $a(L) \leq 1$. A flat Minkowski 5D spacetime corresponds to the choice $a(y)=1$. To keep our discussion general, at this stage we do not specify the warp factor. Choosing different functional forms for $a(y)$ will lead to different (curved) backgrounds.

Once the 5D gauge symmetry is broken down by appropriate boundary conditions to a subgroup of $\mathcal{G}$ on the UV and IR branes, the fifth components of the gauge fields along the broken gauge group generators give rise to scalar excitations with a flat potential at tree level. In a dual description, these scalars can be viewed as 4D Goldstone bosons originated from the spontaneous global symmetry breaking. An effective one-loop potential for the pseudo Nambu-Goldstone boson can be then derived \`{a} la Coleman-Weinberg, starting from the 5D KK theory. One obtains~\cite{Falkowski:2006vi}
\begin{align}
V(\phi) = \sum_r\frac{N_r}{(4\pi)^2}\!\! \int_0^\infty\!\! dp \,p^3\,
\ln\!\left[1+ F_r(-p^2)\sin^2(\lambda_r\,\phi/f)\right],\label{eq:CWpot}
\end{align}
where $f$ is the scale of spontaneous global symmetry breaking, $N_r$ is the number of degrees of freedom of a given particle, $\lambda_r$ is a discrete number that depends on the gauge group structure, and the sum is performed over all zero modes. The quantities $F_r$ are form factors which depend on the specific 5D background. In a 5D setup, these factors are in principle calculable, and for the cases we will be interested in, they are approximately given by the expression~\cite{Falkowski:2006vi}
\begin{align}
F_r(-p^2) \approx \frac{g_r^2 f^2}{M^2_{c} \sinh^2(p/M_{c})}~, \label{eq:formfactor}
\end{align}
which then leads to the effective potential given in Eq.~(\ref{eq:pGpot}). In this framework, the energy scales $f$ and $M_{c}$ are related to 5D geometrical quantities~\cite{Falkowski:2006vi},
\begin{align}
f^{-2} &= g_5^2 \int_0^L dy\,a^{-2}(y)~,\label{eq:f5D}\\
M_{c}^{-1} &= \int_0^L dy\,a^{-1}(y)~,\label{eq:Mkk5D}
\end{align}
where $g_5$ is the 5D gauge coupling which is related to the effective 4D coupling by $g_5 = g_4\sqrt{L}\,$.

To simplify our discussion, in the following we will only consider one zero gauge mode ($r=1$) and identify the corresponding scalar field $\phi$ with the inflaton. Let us start by analyzing the pNGB potential in more detail. We define the quantity (the subscript $r$ is omitted from now on)
\begin{align}
\delta = \frac{g\, f}{M_{c}}~,
\end{align}
and write both $\phi$ and $f$ in units of the 4D Planck mass
\begin{align}
\varphi=\frac{\phi}{M_P}~,\quad \chi= \frac{f}{2\,\lambda\,M_P}~.
\end{align}
In this way, the potential given in Eq.~(\ref{eq:pGpot}) can be rewritten as
\begin{align}
V(\varphi) = \frac{3 M_c^4}{(4\pi)^2} \int_0^\infty d\tilde{p}
\,\tilde{p}^3\, \ln\left[1+\delta^2\,\frac{\sin^2( \varphi/
2\chi)}{\sinh^2(\tilde{p})}\right]~,\label{eq:pGpotII}
\end{align}
with the integration variable $p$ replaced by $\tilde{p}=p/M_c$. It is now easy to see that the compositeness scale $M_c$ is (related to) the scale of the inflationary potential.

Performing the integration, one can see that for $\delta~\leq~1$ the potential can be well approximated by the usual natural cosine form
\begin{align}
V(\varphi)=\Delta^4\left[1-\cos\left(\varphi/\chi\right)\right]~,\label{eq:cospot}
\end{align}
with the potential scale given by
\begin{align}\label{eq:delta1}
\Delta^4 = \frac{3}{(4\pi)^2}\frac{93 \, \zeta(5)}{128}\,M_{c}^4~,
\end{align}
for $\delta=1$ (flat 5D background), and
\begin{align}\label{eq:deltaless1}
\Delta^4 = \frac{3}{(4\pi)^2}\frac{3 \, \zeta(3) }{4}\,\delta^2
\,M_{c}^4~,
\end{align}
for $\delta \ll 1$. For $\delta \gg 1$, the cosine-type potential is not a good approximation, but as it turns out, $\delta \lesssim 1$ is always satisfied for the cases of interest.

\section{Cosine-type natural inflation and WMAP 5-year data}
\label{sec:cosine}

In this section we shall follow the recent analysis of Ref.~\cite{Savage:2006tr} to review the present observational constraints on slow-roll inflation driven by the potential given in Eq.~(\ref{eq:cospot}), taking into account the WMAP5 data~\cite{Hinshaw:2008kr,Dunkley:2008ie,Komatsu:2008hk}. In the next section we will make the connection between the results obtained for this model and natural inflation in different 5D backgrounds.

\subsection{Slow-roll inflation}

The relevant slow-roll inflationary parameters are
\begin{align}
\epsilon(\varphi) &=
\frac{1}{2}\left[\frac{V'(\varphi)}{V(\varphi)}\right]^2
=\frac{1}{2 \chi^2}\left[\frac{\sin(\varphi/\chi)}{1-\cos(\varphi/\chi)}
\right]^2 ~,\\
\eta(\varphi) &=\frac{V''(\varphi)}{V(\varphi)}
=\frac{1}{\chi^2}\left[\frac{\cos(\varphi/\chi)}{1-\cos(\varphi/\chi)}
\right]~,
\end{align}
where the prime denotes derivative with respect to the dimensionless field $\varphi$. In this approximation, inflation ends when the field reaches a value $\varphi_f$ such that one of the slow-roll conditions, $\epsilon < 1$ or $|\eta| < 1$, is violated. From $\epsilon=1$ one obtains
\begin{align}
\cos\left(\varphi_f/\chi\right) = \frac{2\,\chi^2-1}{2\,\chi^2+1}~.
\end{align}

In the slow-roll approximation, the number of e-folds during the inflationary period is given by
\begin{align}
N(\varphi) &\simeq \int_{\varphi_f}^{\varphi}
\frac{V(\bar{\varphi})}{V'(\bar{\varphi})}\,d\bar{\varphi}\nonumber\\
&=\chi^2\left\{\ln\left(\frac{2\,\chi^2}{2\,\chi^2+1}\right)
-2\ln\left[\cos\left(\frac{\varphi}{2\,\chi}\right)\right]\right\}~.
\end{align}

The prediction for the inflationary variables typically depends on the number of e-folds of inflation, $N_\star = N(\varphi=\varphi_\star)$, occurring after the observable Universe leaves the horizon. Although several assumptions about $N_\star$ can be found in the literature, the determination of this quantity requires a model of the entire history of the Universe. While from big bang nucleosynthesis onwards this is now well established, at earlier epochs there are considerable uncertainties such as the mechanism ending inflation and the details of the reheating process. Nevertheless, it is possible to derive a conservative model-independent upper bound: $N_\star<60$~\cite{Liddle:2003as,Dodelson:2003vq}. In fact, $N_\star=55$ is found to be a reasonable fiducial value with an uncertainty of about 5 e-folds around that value.

\begin{figure*}[t]
\begin{center}
\includegraphics[width=7.2cm]{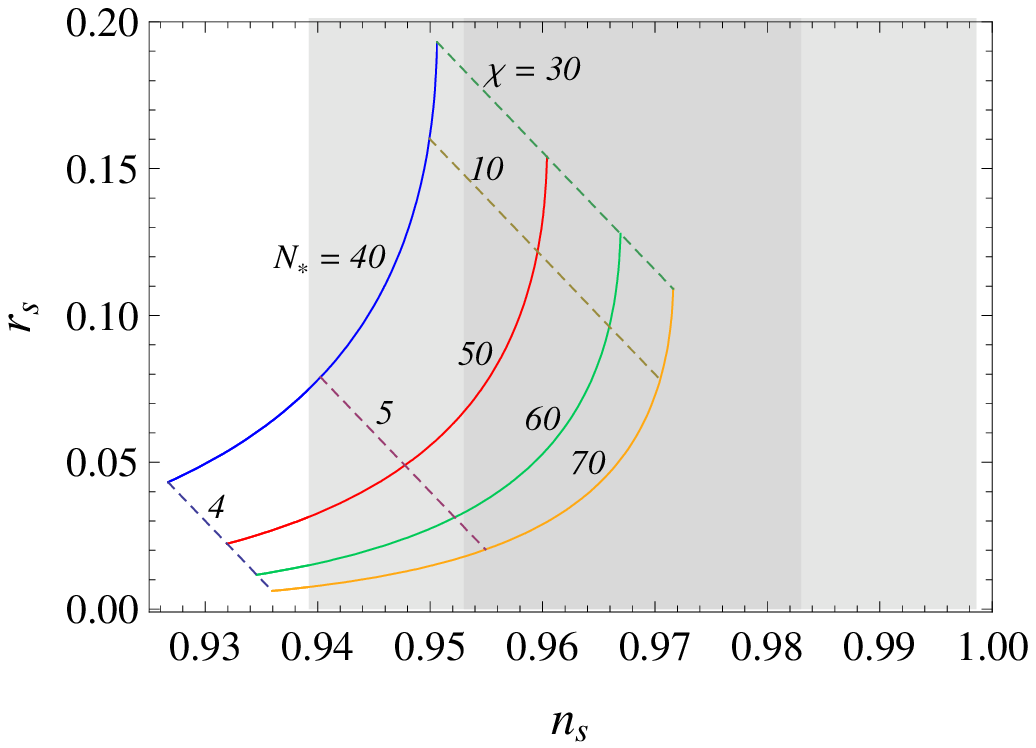}
\includegraphics[width=7.5cm]{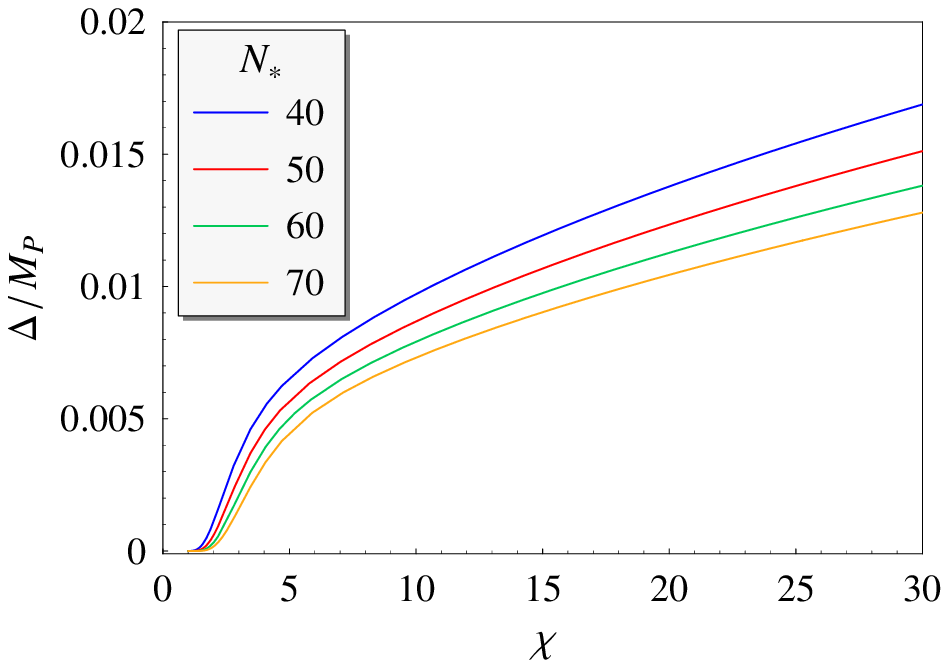}
\end{center}
\caption{(a) Predictions for the cosine-type natural inflation in the $n_s-r_s$ plane. The solid lines are the predictions for
constant $N_\star$ and the dashed lines for constant $\chi$. The shaded areas correspond to the $1\sigma$ and $2\sigma$ intervals
for $n_s$ as given in Eq.~(\ref{eq:nsbound}). (b) The scale $\Delta$ of the potential, in Planck units, as a function of the
spontaneous symmetry breaking scale $\chi$, which is obtained from the WMAP amplitude for the density fluctuations.}
\label{fig:nsrsplane}
\end{figure*}

Indeed, using the slow-roll approximation one can write~\cite{Liddle:2003as}
\begin{align}
N_\star = \ln \frac{k_\star^{-1}}{3000\, h^{-1} \mbox{Mpc}} +
\frac{1}{12}\ln \frac{\rho_{\rm reh}}{\rho_{\rm end}} +
\frac{1}{4}\ln \frac{\rho_{\rm eq}}{\rho_{\rm end}} \nonumber\\
+\ln \frac{H_\star}{H_{\rm eq}} + \ln 217.7\, \Omega_m\, h\,,
\label{eq:reh}
\end{align}
where $H_{\rm eq} = 1.1 \times 10^{-35} h^4\, \Omega_m^2$ and $\rho_{\rm eq}$ are the values of the Hubble radius and the energy
density at the matter-radiation equality epoch, respectively; $H_\star$ is the Hubble radius at the horizon crossing scale
$k_\star\,$; $\rho_{\rm end}$ and $\rho_{\rm reh}$ are the energy density values at the end of inflation and of the reheating
period, respectively. The fractional matter energy density at present is $\Omega_m$. To estimate $N_\star$ we use the WMAP5
central values $h=0.71, ~\Omega_m h^2=0.137$~\cite{Komatsu:2008hk} and the fact that the CMBR anisotropy measured by WMAP allows a determination of the fluctuation amplitude at the scale $k=0.002~\mbox{Mpc}^{-1}$. Taking into account that $(1\,
\mathrm{GeV})^4 \lesssim \rho_{\rm reh} \lesssim \rho_{\rm end}$, one finds that $N_\star$ must lie in the range $48 \lesssim
N_\star \lesssim 61$, for $\chi \gtrsim 4$. One should notice however that there are several ways in which $N_\star$ could lie
outside this range in either direction~\cite{Liddle:2003as,Dodelson:2003vq}. Therefore, in what follows we shall use the range $N_\star = 40 - 70$.

The scalar perturbation amplitude is given by
\begin{align}
A_s^2 = \left.\frac{1}{25 \pi^2}
\frac{H^4}{\dot{\varphi}^2}\right|_{k=k_\star}\!\! = \frac{1}{150 \pi^2
M_P^4}\frac{V(\varphi_\star)}{\epsilon(\varphi_\star)}~,
\label{eq:As}
\end{align}
which for the potential in Eq.~(\ref{eq:cospot}) reads
\begin{align}
A_s^2 =  \frac{\chi^2}{75 \pi^2}\left(\frac{\Delta}{M_P}\right)^4
\frac{\left[1-\cos\left(\varphi_\star/\chi\right)\right]^3}{\sin^2
\left(\varphi_\star/\chi\right)}~.\label{eq:AsII}
\end{align}
For a given $N_\star\,$, the value of the field at horizon crossing can be determined as
\begin{align}
\cos\left(\frac{\varphi_\star}{2\,\chi}\right) = \left(
\frac{2\,\chi^2\,e^{-N_\star/\chi^2}}{1+2\,\chi^2}\right)^{1/2}~.
\end{align}
The spectral tilt for scalar perturbations can be written in terms of the slow-roll parameters as
\begin{align}
n_{s} - 1   \equiv \frac{d\ln A_{s}^2}{d\ln k} =
-6\,\epsilon_\star + 2\,\eta_\star~, \label{eq:ns}
\end{align}
while the tensor power spectrum with amplitude given by
\begin{align}
A_t^2 &= \left.\frac{H^2}{50 \pi^2 M_P^2}\right|_{k=k_\star}\label{eq:At}
\end{align}
can be parameterized in terms of the WMAP normalized~\cite{Spergel:2006hy} tensor-to-scalar ratio as
\begin{align}
r_s\equiv 16 \,\frac{A_t^2}{ A_s^2} = 16\,
\epsilon_\star~.\label{eq:rs}
\end{align}

\subsection{Observational constraints}

The recent publication of the five-year results of WMAP~\cite{Hinshaw:2008kr,Dunkley:2008ie,Komatsu:2008hk} puts very accurate constraints on the spectral index: $n_s = 0.963^{+0.014}_{-0.015}$ at $68\%$ C.L., for vanishing running and no tensor modes.

In what concerns the tensor modes, WMAP5~\cite{Komatsu:2008hk} implies $r_s < 0.43$ (with vanishing running) and $r_s < 0.58$ (with running), both at $95\%$ C.L.. However, models with higher values of $r_s$ require larger values of $n_s$ and lower amplitude of the scalar fluctuations in order to fit the CMBR data, and these are in conflict with large scale structure measurements in the case of vanishing running. If running index is allowed, the large tensor components are still consistent with the data. Hence the strongest overall constraints on the tensor mode contribution comes from the combination of CMBR, large scale structure data and supernovae measurements. The combination of WMAP5, baryon acoustic oscillations (BAO) in the distribution of galaxies, and supernovae (SN) give $r_s < 0.20$ (without running) and $r_s < 0.54$ (with running), at $95\%$ C.L.. If the Lyman-$\alpha$ forest spectrum from Sloan Digital Sky Survey is also considered~\cite{Seljak:2006bg}, then~\cite{Komatsu:2008hk} $r_s < 0.28$ (at $95\%$ C.L. and with running).

\begin{figure*}[tb]
\begin{center}
\includegraphics[width=7.5cm]{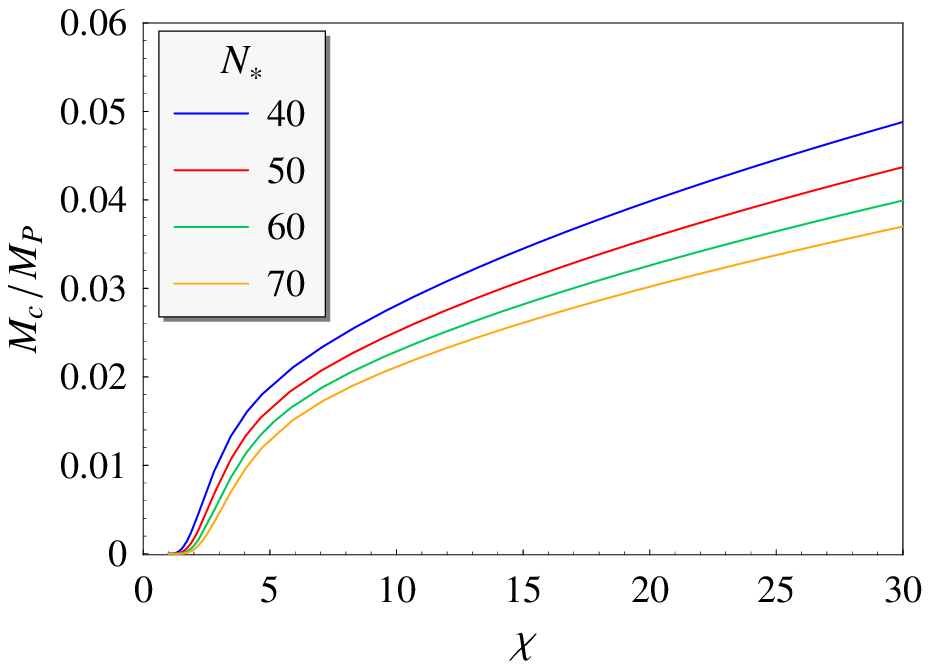}
\includegraphics[width=7.5cm]{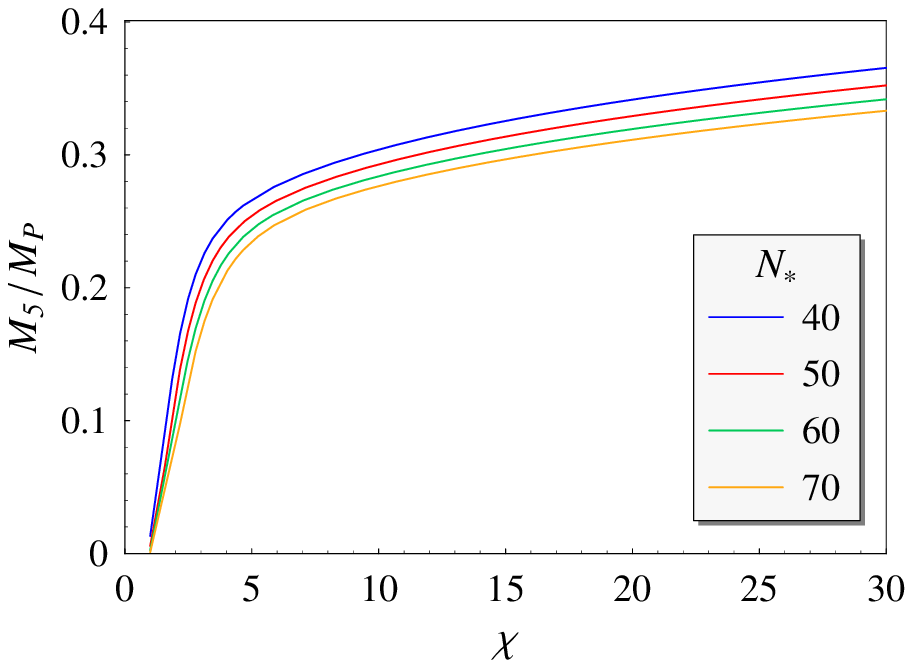}\\
\end{center}
\caption{The mass scales $M_c$ and $M_5$ in a 5D flat background as functions of the spontaneous symmetry breaking parameter $\chi$.} \label{fig:flat}
\end{figure*}

Since the running $\alpha_s$ is very small in the natural inflation model under consideration, $\alpha_s \sim 10^{-3}$, we can make use of the observational bounds obtained for the case of vanishing running. However, one has to take into account the tensor modes and their inclusion has implications for $n_s$. With the combined data (WMAP5+BAO+SN) the constraints become~\cite{Komatsu:2008hk}
\begin{align}\label{eq:nsbound}
0.953 &< n_s < 0.983~(\text{at 68\% C.L.})~,\nonumber\\
0.9392 &< n_s < 0.9986~(\text{at 95\% C.L.})~.\end{align}
These are the bounds we will consider in our analysis.

In Fig.~\ref{fig:nsrsplane} we present the predictions for the cosine-type natural inflation in the $n_s-r_s$ plane. The solid and dashed lines are curves for constant $N_\star$ and $\chi$, respectively. The darker (lighter) gray shaded area corresponds to the $68\%$ ($95\%$) C.L. for $n_s$ given in Eq.~(\ref{eq:nsbound}). In general, both $n_s$ and $r_s$ depend on $\chi$ and $N_\star$:
\begin{align}
n_s &= 1 - \frac{1}{\chi^2} - \frac{4}{(1 + 2\,\chi^2)\exp(N_\star/\chi^2)
-2\,\chi^2}~,\\
r_s &= \frac{16}{(1 + 2\,\chi^2)\exp(N_\star/\chi^2)-2\,\chi^2}~,
\end{align}
but one can easily find that~\cite{Savage:2006tr}
\begin{align}
n_s \approx \left\{\begin{array}{c}
  1-\frac{1}{\chi^2}\quad {\rm for~}\chi \ll 1~, \\
  1- \frac{2}{N_\star}\quad {\rm for~}\chi \gg 1~,\\
\end{array}\right.
\end{align}
and
\begin{align}
r_s \approx \left\{\begin{array}{c}
  0\quad {\rm for~}\chi \ll 1~, \\
  \frac{8}{N_\star}\quad {\rm for~}\chi \gg 1~.\\
\end{array}\right.
\end{align}
Hence, for sufficiently large $\chi$ the behavior is similar to the quadratic potential,  since inflation occurs near the minimum of the potential, where it can be approximated by a parabola~\cite{Savage:2006tr}.

From the observational constraint in the spectral index $n_s$ we conclude that the minimum value allowed for the scale of spontaneous global symmetry breaking is [cf. Figure~\ref{fig:nsrsplane}(a)]
\begin{align}\label{eq:chimin2s}
    \chi_{\rm min} \simeq 4.12~,\quad f_{\rm min} \simeq 8.24\,\,\lambda_r\, M_P~,
\end{align}
at $95\%$ C.L. and
\begin{align}\label{eq:chimin1s}
    \chi_{\rm min} \simeq 4.85~,\quad f_{\rm min} \simeq 9.70\,\,\lambda_r\, M_P~,
\end{align}
at $68\%$ C.L..
Thus, as mentioned in the introduction, one must have super-Planck values for the scale $f$ in order to comply with the observational data requirements on $n_s$. Nevertheless, in an extra-dimensional setup this is not necessarily problematic, since the effects of higher-dimensional operators can be kept under control and locality in extra dimensions can prevent the inflaton potential from acquiring large corrections~\cite{Arkani-Hamed:2003mz}. In our subsequent analysis we will take $\chi \gtrsim 4$ as the lower bound for the scale of spontaneous symmetry breaking.

The scale $\Delta$ of the potential can be determined from Eq.~(\ref{eq:As}) by imposing the correct amplitude for the density fluctuations $A_s=A_s^{\rm cmb}$, as measured by the WMAP team: $A_s^2(k=0.002~ {\rm Mpc}^{-1}) \approx 4 \times 10^{-10}$~\cite{Spergel:2006hy,Komatsu:2008hk}. The allowed values for $\Delta$ are shown in Fig.~\ref{fig:nsrsplane}(b), where we can see that a potential height of the order of $\Delta \gtrsim 5 \times 10^{-3} M_P$ is required.

Before addressing the connection of the results presented above with the extra-dimensional quantities in different spacetime backgrounds, we should remark that, for $\chi \gtrsim 4$, it is possible to obtain a sufficiently large total number of e-folds of inflation, $N_{tot} \gtrsim 70$, as required in order to solve the initial condition problems of standard cosmology.

\section{Natural inflation in a flat 5D spacetime}

We begin by considering the simplest theory with one extra dimension: a 5D gauge theory in a flat Minkowski background. In this case the warp factor $a(y)=1$ and Eqs.~(\ref{eq:f5D}) and (\ref{eq:Mkk5D}) imply the mass scale relations
\begin{align}
 g\,f = M_{c} = L^{-1}\,.
\end{align}
Moreover, the 4D and 5D fundamental Planck masses are simply related by~\cite{Arkani-Hamed:1998rs,Arkani-Hamed:1998nn}
\begin{align}
M_P^2=M_5^3\, L\,.
\end{align}
The generated pNGB inflaton potential is determined by Eqs.~(\ref{eq:cospot}) and (\ref{eq:delta1}),
\begin{align}
V(\varphi)&=\frac{279 \, \zeta(5)}{2048\,\pi^2} \,M_{c}^4
\left[1-\cos\left(\varphi/\chi\right)\right] \nonumber\\
&\simeq \frac{9}{64\,\pi^2L^4}
\left[1-\cos\left(\varphi/\chi\right)\right]~.\label{eq:cospotflat}
\end{align}

\begin{figure}[tb]
\begin{center}
\includegraphics[width=8cm]{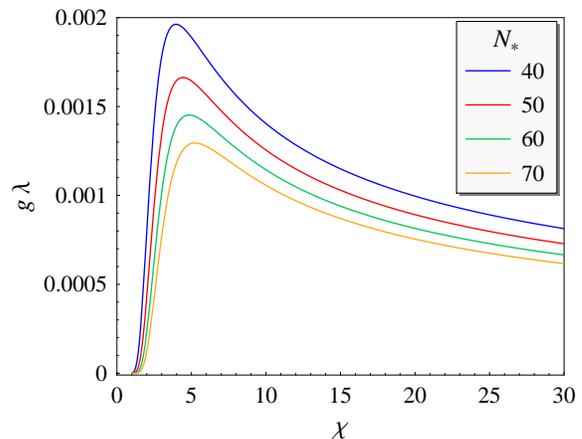}\\
\end{center}
\caption{The 4D effective coupling $g\lambda$ as a function of $\chi$, in a 5D flat background.} \label{fig:flat2}
\end{figure}

The inflationary predictions for this scenario are similar to the so-called extra-natural inflation model of
Ref.~\cite{Arkani-Hamed:2003wu} (see also \cite{Kaplan:2003aj,Arkani-Hamed:2003mz}), where the inflaton is identified with the phase $\theta$ of the gauge-invariant Wilson loop of the fifth component $A_5$ of an Abelian gauge field propagating in the bulk, i.e. $\theta=g_5\oint dy\,A_5\,$. In the presence of particles charged under the Abelian symmetry, and at energies below $1/R$ ($R$ is the compactification radius), the canonically normalized 4D field $\phi=\theta f$ develops an effective one-loop potential, which in leading order has the form~(\ref{eq:cospotflat}) with the effective decay constant of the spontaneously broken Abelian symmetry given by $f = 1/(2\pi\,R\, g)$.

Normalizing to the WMAP anisotropy measurements for the curvature density, it is possible to determine the mass scales $M_c$ and
$M_5$ as functions of the symmetry breaking parameter $\chi$ (see Fig.~\ref{fig:flat}). From the lower bound $\chi \gtrsim 4$ we
obtain the constraint
\begin{align}\label{eq:lowerMcflat}
M_c \gtrsim 2.3 \times 10^{16}\, {\rm GeV} = 9.6 \times 10^{-3}\,M_P~,
\end{align}
i.e. the size of the extra dimension should be very small, $L \lesssim 4.4 \times 10^{-17}\,{\rm GeV}^{-1}  \simeq
10^{2}\,M_P^{-1}$. This in turn implies the following lower bound on the fundamental 5D Planck mass $M_5$:
\begin{align}\label{eq:lowerM5flat}
M_5 = (M_c\, M_P^2)^{1/3} \gtrsim 5.1 \times 10^{17}\, {\rm GeV}= 0.2 \,M_P~.
\end{align}
Therefore, quantum gravity  corrections to the inflaton potential are expected to be negligible because the Planck length is much
smaller than the size of the extra dimension. Furthermore, during the inflationary period the Universe can be considered as four dimensional, since the Hubble length $H^{-1} \sim M_P/V^{1/2}$ is larger than the size of the extra dimension. Finally, we should remark that due to the smallness of the inflationary parameters, a small value of the effective 4D gauge coupling $g$ is always required in this framework, as illustrated in Fig.~\ref{fig:flat2}.

\section{Natural inflation in warped backgrounds}

\begin{figure}[b]
\begin{center}
\includegraphics[width=8cm]{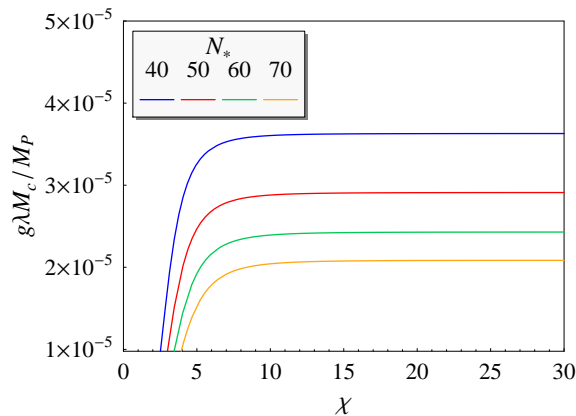}\\
\end{center}
\caption{The inflationary predictions for the parameter $g \lambda M_c$ (in Planck units) as a function of $\chi$, in a generic 5D curved background for $g f \lesssim M_c\,$.}
\label{fig:curved}
\end{figure}

\begin{figure*}[tb]
\begin{center}
\includegraphics[width=7.5cm]{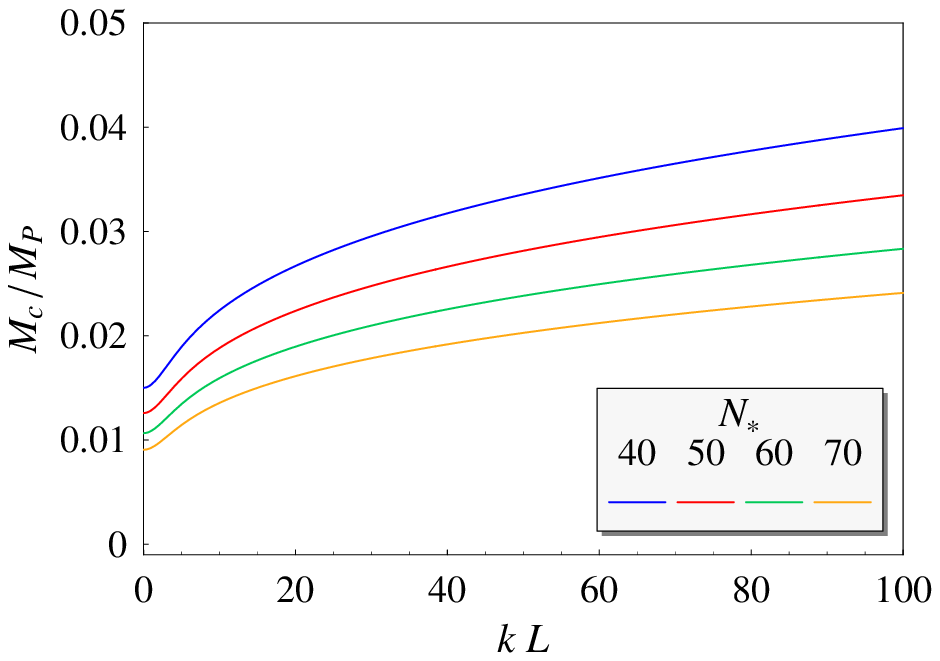}
\includegraphics[width=7.5cm]{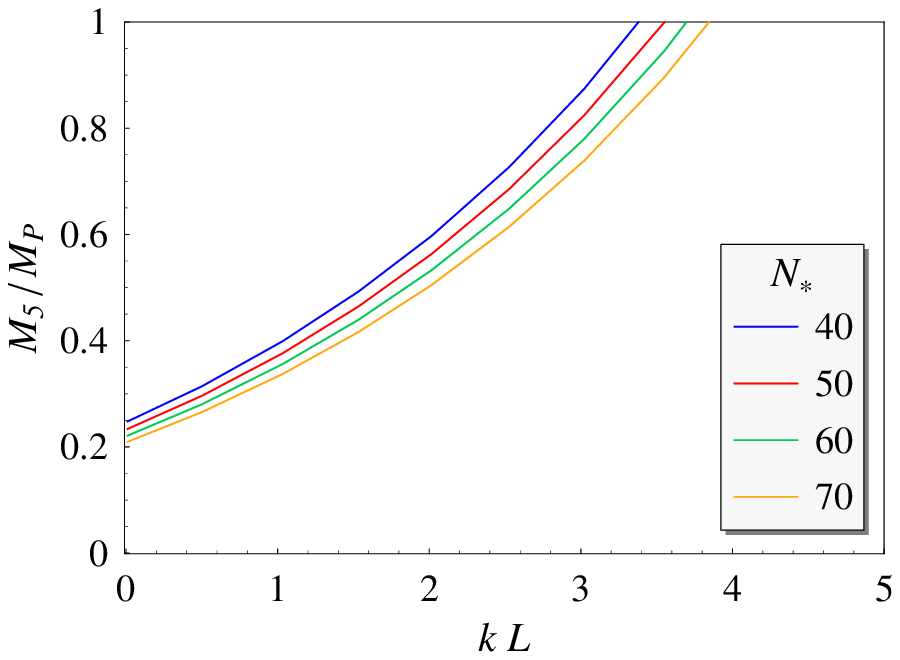}\\
\end{center}
\caption{Lower bounds on $M_c$ and $M_5$, in Planck units, for the RSI 5D background and taking $\chi=4$.} \label{fig:RSI}
\end{figure*}

Warped extra dimensions provide a simple geometrical picture in which it is possible to address some of the problems of the standard model (e.g. the hierarchy problem and electroweak symmetry breaking). But even more compelling is the fact that this alternative 5D framework admits a 4D holographic interpretation in terms of a strongly coupled gauge theory~\cite{ArkaniHamed:2000ds,Rattazzi:2000hs,PerezVictoria:2001pa}. Once again we assume that inflation is driven by the pNGB inflaton potential predicted by the 5D gauge theory. For a given warp factor $a(y) \neq 1$, the relevant inflationary energy scales $f$ and $M_c$ are related to the 5D geometry through Eqs.~(\ref{eq:f5D}) and (\ref{eq:Mkk5D}). For $g f < M_c$ the potential is well approximated by Eqs.~(\ref{eq:cospot}) and (\ref{eq:deltaless1}), which yield
\begin{align}
V(\varphi)=\frac{9\, \zeta(3)}{16\,\pi^2}\,(g \lambda M_{c})^2 M_P^2 \chi^2
\left[1-\cos\left(\varphi/\chi\right)\right]~.\label{eq:cospotwarped}
\end{align}

From the slow-roll inflationary constraints it is then possible to constrain the combination $g \lambda M_c$, independently of the 5D gravitational background. The results are presented in Fig.~\ref{fig:curved}, where this parameter is plotted as a function of $\chi$ for different e-folds of inflation. We obtain the lower bound
\begin{align}
g\,\lambda\,M_c \gtrsim 2.5 \times 10^{13}\, {\rm GeV}~,
\end{align}
for a successful natural inflation.

Clearly, to fully determine the allowed values of the effective coupling $g$ and the energy scale $M_c$, the warp factor must be specified. Below we shall consider two simple cases: anti-de Sitter and dilatonic spacetime.

\subsection{AdS$_5$ spacetime}

Let us consider the case where the fifth dimension is a slice of AdS$_5$, which corresponds to the so-called Randall-Sundrum I (RSI) model~\cite{Randall:1999ee}. As is well known, the holographic correspondence between the 5D warp model in a slice of AdS$_5$ and the 4D theory originates from the AdS/CFT correspondence in string theory~\cite{Maldacena:1997re,Witten:1998qj,Gubser:1998bc}. In this model the warp factor is given by
\begin{align}\label{warpAdS5}
a(y)=e^{- k\, y}~,
\end{align}
where $k$ is the energy scale corresponding to the curvature $\ell$ of AdS$_5$, which is related to the negative bulk
cosmological constant by
\begin{align}
\Lambda_5=-6\,k^2=-\frac{6}{\ell^2}~.
\end{align}
The two branes localized at $y=0$ and $y=L$ have opposite tensions, $\pm \sigma$, with
\begin{align}
\sigma=\frac{3 M_P^2}{4 \pi\,\ell^2}~.
\end{align}
The fundamental scale $M_5$ of the positive brane is related to the 4D Planck mass by
\begin{align}\label{eq:M5AdS5}
M_P^2=M_5^3 \int_0^L dy\,a^2(y)=\frac{M_5^3}{2\,k}\,(1-a_L^2)~,
\end{align}
where $a_L \equiv a(y=L)=\exp(-k\,L)$. Furthermore, from Eqs.~(\ref{eq:f5D}) and (\ref{eq:Mkk5D}) one obtains~\cite{Falkowski:2006vi}
\begin{align}
f^2 &= \frac{2\,k}{g^2\,L\,(a_L^{-2}-1)}~,\label{eq:fAdS5}\\
M_{c} &= \frac{k}{a_L^{-1}-1}~.\label{eq:MkkAdS5}
\end{align}

\begin{figure}[b]
\begin{center}
\includegraphics[width=8cm]{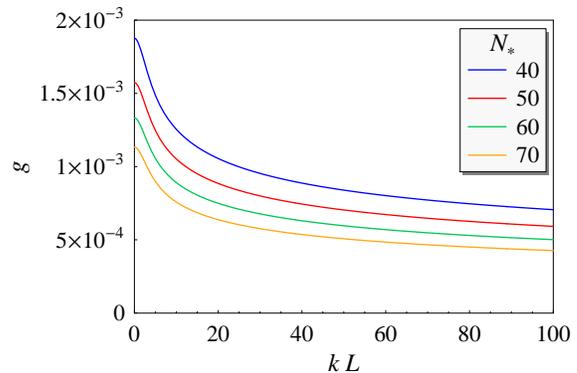}\\
\end{center}
\caption{Upper bound on the 4D effective gauge coupling $g$ for the RSI 5D background. We assume the lower bound $\chi=4$ imposed by natural inflation.} \label{fig:RSI2}
\end{figure}

It is easily verified that $\delta \equiv gf/M_c < 1$ for this background and, therefore, the pNGB potential is given by Eq.~(\ref{eq:cospotwarped}). Using then Eqs.~(\ref{eq:fAdS5}) and (\ref{eq:MkkAdS5}), the inflationary constraints can be expressed in terms of the 5D quantities $k$ and $L$. In Fig.~\ref{fig:RSI} we present the lower bounds on the energy scales $M_c$ (left panel) and $M_5$ (right panel), in Planck units, as functions of $kL$, assuming the lower bound $\chi=4$. Requiring $M_5<M_P$ imposes an upper bound on the warp exponential factor, namely, $kL \lesssim 4$. Clearly, the scales $M_c$ and $M_5$ are bounded from below by the flat bounds given in Eqs.~(\ref{eq:lowerMcflat}) and (\ref{eq:lowerM5flat}), respectively. As in the flat case, the 4D effective coupling is small: $g \sim 10^{-3}$ (cf. Fig.~\ref{fig:RSI2}).

\begin{figure*}[t]
\begin{center}
\includegraphics[width=7.5cm]{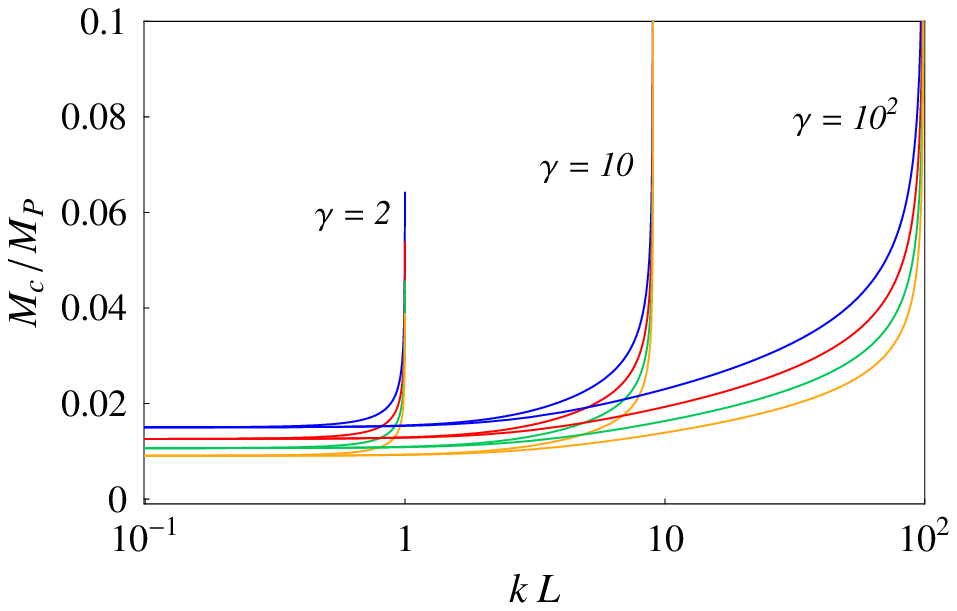}
\includegraphics[width=7.5cm]{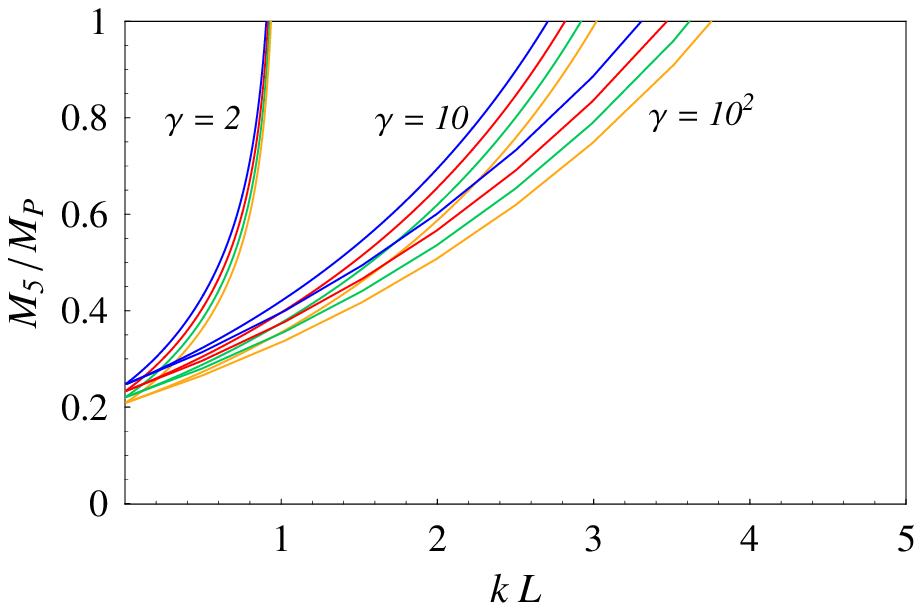}\\
\end{center}
\caption{Lower bounds on $M_c$ and $M_5$, in Planck units,  for the 5D dilatonic background and taking $\chi=4$. The results are
presented for different values of $\gamma$. The different color lines correspond to $N_\ast =40,50,60,70$.} \label{fig:dilatonic}
\end{figure*}

\subsection{Dilatonic spacetime}

We consider a 5D gravity-dilaton theory with the Liouville-type potential~\cite{Chamblin:1999ya,Kachru:2000hf}
\begin{align}
U(\psi) = -6\,k^2\,\frac{\gamma\,(\gamma-1/4)}{(\gamma-1)^2}\,
\exp\left[\pm\frac{2}{\sqrt{3\,\gamma}}\,(\psi-\psi_0)\right]~,
\end{align}
and a potential of the form $\hat{U}(\psi)=\hat{U}_0\exp(\beta\,\psi)$, properly chosen at the domain wall. The solution of the bulk equations of motion, with the metric given in Eq.~(\ref{eq:Mkk5D}) and the ansatz $a(y)=\exp(\alpha\,\psi)$, can then be obtained in the form~\cite{Kachru:2000hf}
\begin{align}
\psi=-\sqrt{3\gamma}\,\ln\left(1-\frac{ky}{\gamma-1}\right)\,, \quad\alpha=-\sqrt{\frac{\gamma}{3}}\,.
\end{align}
The corresponding warp factor of the background metric is therefore given by
\begin{align}
a(y)=e^{\alpha\, \psi}=\left(1-\frac{k\,y}{\gamma-1}\right)^\gamma~.
\end{align}
The parameter $\gamma$ is a measure of the explicit breaking of the conformal symmetry in the dual 4D strongly coupled theory. In the conformal limit, i.e. when $\gamma \rightarrow \infty$, the AdS$_5$ exponential warp factor of Eq.~(\ref{warpAdS5}) is recovered. The case $\gamma=1/6$ corresponds to the Ho\v{r}ava-Witten model compactified to five dimensions~\cite{Horava:1996ma, Lukas:1998tt, Lukas:1998qs, Lukas:1998yy}.

For this dilatonic spacetime we find the relation
\begin{align}
M_P^2 = M_5^3 \int_0^L dy\,a^2(y) = \frac{M_5^3\,(\gamma-1)}{k\,(2\,\gamma+1)} \left(1-a_L^{2+1/\gamma}\right)\,, \label{eq:M5dil}
\end{align}
while from Eqs.~(\ref{eq:f5D}) and (\ref{eq:Mkk5D}) one obtains~\cite{Falkowski:2006vi}
\begin{align}
f^2 &=\frac{(2\gamma-1)\,k}{g^2\,(\gamma-1)\,L\, \left(a_L^{-2+1/\gamma}-1\right)}~,\label{eq:fdil}\\
M_{c} &= \frac{k}{a_L^{-1+1/\gamma}-1}~.\label{eq:Mkkdil}
\end{align}
In the conformal limit the above expressions reproduce Eqs.~(\ref{eq:M5AdS5})-(\ref{eq:MkkAdS5}). It is also
straightforward to check that $\delta \equiv g\, f/M_{c}<1$ and the pNGB inflaton potential is simply given by Eq.~(\ref{eq:cospotwarped}).

\begin{figure}[b]
\begin{center}
\includegraphics[width=8cm]{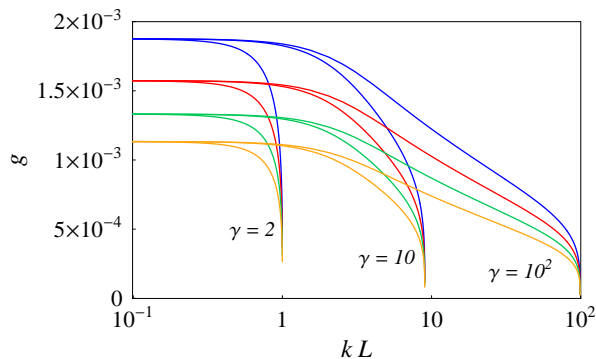}\\
\end{center}
\caption{Upper bound on $g$, for a 5D dilatonic background and different values of $\gamma$. We assume $\chi \gtrsim 4$, as
imposed by the inflationary constraints. } \label{fig:dilatonic2}
\end{figure}

In Fig.~\ref{fig:dilatonic} we present the predictions for  the mass scales $M_c$ and $M_5$ as functions of the warp exponent $kL$ and for different values of $\gamma$. As in the RSI case, these scales are bounded from below by the flat bounds given in
Eqs.~(\ref{eq:lowerMcflat}) and (\ref{eq:lowerM5flat}). Moreover, the upper bound on the 4D effective coupling is $g \sim
\mathcal{O}(10^{-3})$, as shown in Fig.~\ref{fig:dilatonic2}.

\section{Conclusions}

During the last few years there has been considerable activity on various aspects of braneworld cosmology. Branes and warped extra dimensions might have played an important role in early Universe cosmology. In particular, they might be relevant in solving some of the problems of low-energy effective field theories, such as the mass hierarchy and symmetry breaking issues. In this paper we have addressed the question of how a pseudo Nambu-Goldstone potential, typically predicted in 5D gauge theories, can lead in a natural way to a successful inflation in the four-dimensional imprint of the brane world.

The fact that the 4D effective coupling is very small in all the cases considered, $g \lesssim \mathcal{O}(10^{-3})$, seems
unavoidable in the natural inflation scenario. This is a direct consequence of the smallness of the inflationary parameters and
density fluctuations. We should also remark that the bounds imposed by natural inflation on the warped fifth dimension are quite restrictive. For AdS$_5$ and dilatonic gravitational backgrounds we have found the constraint $kL \lesssim 4$, an upper bound which is too small to naturally generate a TeV mass scale at the IR brane.

As mentioned before, the predicted running of the spectral index is negligible in the models we have considered, and too small to be detected in the forthcoming small scale CMB experiments. However, one should notice that larger values of $\alpha_s$ can
be obtained if we consider additional zero modes in the inflaton potential. For instance, in the case where both a bosonic and a
fermionic zero mode contribute to the pNGB potential, an inflaton potential of the form
\begin{align}
V(\varphi)=\Delta^4 \left[1 -\rho - \cos\left(\frac{\varphi}{\chi}\right) + \rho
\cos\left(\frac{\kappa\,\varphi}{\chi}\right)\right]~,
\end{align}
is obtained, where $\chi \equiv \chi_b$, $\Delta \equiv \Delta_b$, $\kappa \equiv \chi_b/\chi_f=\lambda_f/\lambda_b$, and
$\rho \equiv \Delta_f^4/\Delta_b^4= 4 \,y_f^2/(3 \,g_b^2) $. The subscripts $b$ and $f$ refer to the bosonic and fermionic
quantities, respectively. Considering the regime where $\rho \ll 1$, $\kappa \gg 1$, $\rho\, \kappa \ll 1$, and $\rho\,\kappa^2 \sim \mathcal{O}(1)$, it is possible to achieve $|\alpha_s| \sim \mathcal{O}(10^{-2})$~\cite{Feng:2003mk}, as favored by current WMAP results.

\begin{acknowledgments}
N.M.C.S. acknowledges the support of the \emph{Funda\c{c}\~{a}o para a Ci\^{e}ncia e a Tecnologia} (FCT, Portugal) through Grant No. SFRH/BPD/36303/2007.
\end{acknowledgments}


\end{document}